\documentclass[runningheads]{llncs}
\usepackage{graphicx}
\usepackage{listings}
\usepackage{makecell}
\usepackage{wrapfig}
\begin{document}
\title{The role of interactive super-computing in using HPC for urgent decision making}
\titlerunning{Role of interactive super-computing using HPC for urgent decision making}
%
\author{Nick Brown \inst{1} \and Rupert Nash \inst{1} \and Gordon Gibb \inst{1} \and Bianca Prodan \inst{1} \and Max Kontak \inst{2} \and Vyacheslav Olshevsky \inst{3} \and Wei Der Chien \inst{3}}
\authorrunning{N. Brown et al.}
%
\institute{EPCC, The University of Edinburgh, Bayes Centre, Edinburgh, UK \\
\email{n.brown@epcc.ed.ac.uk}
\and
German Aerospace Center, High-Performance Computing Group, Cologne
\and
KTH Royal Institute of Technology, Stockholm, Sweden}
\maketitle              
\begin{abstract}
Technological advances are creating exciting new opportunities that have the potential to move HPC well beyond traditional computational workloads. In this paper we focus on the potential for HPC to be instrumental in responding to disasters such as wildfires, hurricanes, extreme flooding, earthquakes, tsunamis, winter weather conditions, and accidents. Driven by the VESTEC EU funded H2020 project, our research looks to prove HPC as a tool not only capable of simulating disasters once they have happened, but also one which is able to operate in a responsive mode, supporting disaster response teams making urgent decisions in real-time. Whilst this has the potential to revolutionise disaster response, it requires the ability to drive HPC interactively, both from the user's perspective and also based upon the arrival of data. As such interactivity is a critical component in enabling HPC to be exploited in the role of supporting disaster response teams so that urgent decision makers can make the correct decision first time, every time.

\end{abstract}
\section{Introduction}
\label{sec:Intro}
The ability to perform faster than real-time simulations of large-scale situations is becoming a reality due to the technological advances that computing, and specifically HPC, has enjoyed over the past decade. This opens up a host of opportunities including the ability to leverage HPC in the role of time-critical decision support for unfolding emergency scenarios. However to do this one must extend the state of the art in numerous fields such as in-situ HPC data analytics, the assimilation and ingestion of data sources (e.g. from sensors measuring actual situational conditions), data reduction and statistical sampling for real-time visualization.

The Visual Exploration and Sampling Toolkit for Extreme Computing (VESTEC) project is exploring the fusion of HPC with real-time data for supporting urgent decision makers in the role of disaster response. In this project we are concerned with three use-cases that drive our research in this area. The first one is the spread of mosquito borne diseases, which is a major health challenge around many parts of the world. This spread is heavily influenced by weather and current mosquito conditions, so being able to combine real world mosquito reports, accurate temperature values and short range weather forecasts is critical when simulating how the spread will progress in the short to medium term. 

The second VESTEC use-case is the progression of forest fires and the ability to explore the impact of different response scenarios. Based on the successful Wildfire Analyst application, this code is currently fed manually with data from the field, but this is a labour intensive process generating extra work for the disaster response teams and inducing a time lag on information, limiting the real-time use of this tool. By automating the process it will make this things far more dynamic, reduce the load on the operators and enable a much greater quantity of data to be exploited in the simulation of wildfires. The third use-case is that of space weather, for instance solar storms, which is very costly because of the damage done to satellites. Whilst it is possible for satellites to be switched off, thus protecting them, one needs prior warning and-so being able to ingest the current space weather conditions and run this through simulation tools is a very valuable proposition as it enables the estimation of risk to assets.

Whilst these three use-cases represent very different areas, all with novel challenges, there are a number of general similarities. All three involve the ingestion of large volumes of data and, as it currently stands, all generate large volumes of output data which is then processed offline by downstream tools. All use-cases involve coupled simulations where a number of distinct codes must be executed in a specific order, for example the execution of a high resolution weather model before feeding the results into the mosquito borne diseases simulation. Lastly, all models involve, to some extent, ensembles which need to be rapidly started and stopped based on new data arriving unpredictably, and also steered by the urgent decision maker.

In order to fully leverage HPC for disaster response, one must embrace the interactive nature of these workflows, selecting the appropriate tools and techniques to support how these technologies can support emergency decision makers in their work. The rest of this paper is organised as follows, after briefly considering related work and background in Section \ref{sec:bg}, we then explore the role of interactivity in Section \ref{sec:role} by considering the three main interactivity challenges one faces when using HPC for urgent decision making. In Section \ref{sec:cloud} we explore the role of cloud computing in urgent decision making before drawing conclusions in Section \ref{sec:conclusions}.

\section{Background and Related work}
\label{sec:bg}

\begin{figure}
\centering
\includegraphics[scale=0.8]{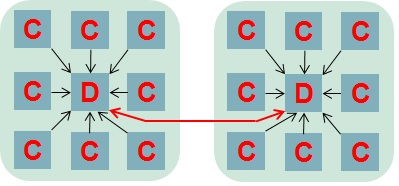}
\caption{In-situ data analytics, where cores are shared between computation and data handling}
\label{fig:dataanalytics}
\end{figure}

In Section \ref{sec:Intro} we highlighted the fact that the three VESTEC use-cases currently write their data to disk for offline processing at some later point. When moving to exploiting HPC in a more real-time urgent decision responsive mode, where operators need the results ASAP, then the overhead of IO makes this offline analytics approach prohibitive. Therefore a logical starting point is to consider the current state of the art in processing data from simulations in-situ. This is illustrated in Figure \ref{fig:dataanalytics}, where cores of a processor are shared between compute (C) and data analytics (D). The general idea is that a number of compute cores are serviced by a data analytics core, and these compute cores \emph{fire and forget} their raw data over to the corresponding data core for analytics whilst the computation proceeds. The major benefit of this is that the raw data is never written to disk, thus avoiding the overhead of IO and the compute cores can continue to work with data analytics proceeding concurrently. Typically this data analytics involves some form of data reduction, or live visualisation.

Existing frameworks such as XIOS \cite{xios}, Damaris \cite{damaris}, ADIOS \cite{adios}, and MONC \cite{monc} are examples of in-situ data analytics commonly used by HPC simulation codes. While these approaches have proven effective and efficient at interleaving data analytics with running simulations none of them do so in real-time, address the issues of ensemble simulation or, most importantly, support the assimilation and incorporation of new data into running simulations. These limitations are key requirements for a system that is required to support urgent decision making and, as such, one needs to go further than the current state of the art in in-situ data analytics.

More generally there have been previous attempts at using HPC for urgent decision making. One such example was the SPRUCE project \cite{spruce}, which agreed with machine owners a priori that their resource could be used to run urgent workloads. Responders were given tokens which represented the amount of time they had available on the specific machine for their workload. The solution was somewhat more nuanced than this because a number of modes of access were supported. These ranged from jobs submitted with some sort of priority in the batch queue, all the way to forcibly terminating already running jobs and replacing them for with the urgent job. The more severe the mode, the more user tokens were used up.

However these previous uses of HPC for urgent decision making still involved a batch processing model, where operators submit jobs, commonly with some priority in the queue, and once these have completed the results are used to inform response decisions. However this is not sufficient for our approach because, to take advantage of the high velocity data and live data analytics methods which are becoming commonplace, a much more interactive pattern must be embraced. The traditional batch processing organisation, which has served classic HPC workloads so well over many decades, is simply not appropriate here as we require the execution of ensemble models driven by the unpredictable arrival of data and chaining of HPC codes. This is further complicated by the fact that, in order for the results to be of use to the disaster response team, these urgent jobs must run within a specific bounded time. Furthermore, as simulations are then progressing it is crucially important for a user to be able to interacting with them and exploring numerous options and response techniques.

\section{The role of interactivity}
\label{sec:role}
We believe that using HPC for urgent decision making requires interactivity in three main areas:
\begin{enumerate}
    \item \textbf{User interaction with running simulations}: gaining feedback as the code is running and the ability to modify the simulation state and parameters as it is executing
    \item \textbf{Dynamic ensemble simulations}: where jobs can be started and stopped on the fly. This is either driven directly by the user, or automatically based on the arrival of sensor data
    \item \textbf{Supporting an interactive workload}: where the arrival of new data, or completion of another code, such as a weather model, automatically starts new jobs
\end{enumerate}

\begin{figure}
\centering
\includegraphics[scale=0.6]{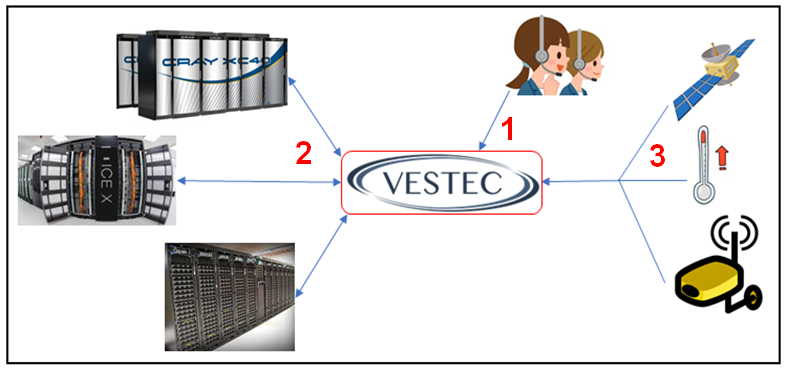}
\caption{Overview of interactivity involved in using HPC for urgent decision making}
\label{fig:interactivity}
\end{figure}

These three points are illustrated in Figure \ref{fig:interactivity}, where a central VESTEC system connects data sources and disaster response teams, to HPC machines. Taking the Wildfire Analysist use-case as an example, it was highlighted in Section \ref{sec:bg} that until now interaction with this code is manual, where first line responders have to explicitly start new simulations based upon updated information they have received and entered. Instead, as per Figure \ref{fig:interactivity}, the ability to drive the code dynamically, so that data is automatically picked up as it streams in and leveraged according to some predefined rules, relieves considerable pressure from the disaster response team and potentially results in much more accurate, up to date, perspective of the situation.

In this section We explore the challenges and solutions associated with each of these three components of interactivity in using HPC for urgent decision making.

\subsection{Interacting with running simulations}
At many GBs in size, the data generated by the codes involved in our three use-cases is substantial and, as discussed in Section \ref{sec:bg}, it is already understood that, to enable efficient data processing, then an in-situ method, where the data is handled as it is generated rather than written out to disk first, is crucial. Furthermore, there is a requirement for feeding back from the user to the simulation in order to modify the state \emph{on the fly} as the code is running. The unique requirements of urgent decision making means that driving this analysis and interaction via a visual program is highly desirable and ParaView \cite{paraview} is being used for this in the VESTEC project.

ParaView \cite{paraview} is an open source data analysis and visualisation application used in many different application areas to analyze and visualize scientific data sets. Designed as a framework, tools are provided to build visualisations appropriate to specific application data analytics and then exploration can be performed interactively in 3D or via ParaView's batch processing capabilities. ParaView supports execution over distributed memory and, as such, can handle very large datasets, which is important in the context of disaster response. Under the hood, ParaView uses the highly popular Visualization ToolKit (VTK) for graphics rendering and Qt for windowing support. In this project we are combining ParaView with Catalyst \cite{catalyst}, an in-situ library which orchestrates the simulation with analysis and/or visualization and connects to ParaView. A major benefit of these tools is that the analysis and visualization tasks can be implemented in a high level language such as Python or C++, or via the ParaView GUI.

\begin{figure}
\centering
\includegraphics[scale=0.8]{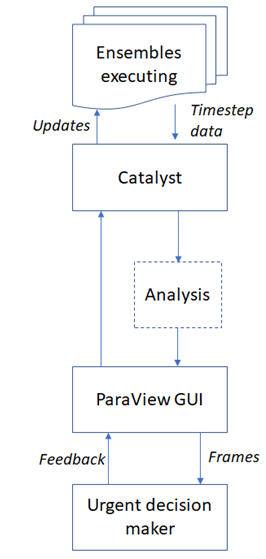}
\caption{Example of the interaction with running ensembles}
\label{fig:example}
\end{figure}

The role of ParaView and Catalyst is illustrated in Figure \ref{fig:example}, servicing a number of ensemble simulations that are running on an HPC machine. These ensembles are integrated with Catalyst and raw timestep data from each job is passed to a user provided pipeline of analysis scripts which reduce it, for instance via sampling. This reduced data is then transmitted to the ParaView GUI, typically running on a client machine and connected by TCP/IP, but it can also pick up files using the file system or run directly on the HPC machine and be forwarded via X11. Depending on how the ParaView GUI has been set up for the specific disaster response job, the user is able to provide feedback which, via the Catalyst tool, is delivered to the appropriate ensemble job(s). Whilst the simulation itself still needs to implement the specific details of how to process the user's feedback, using the Catalyst tool means that they do not need to be concerned with the mechanism of how the feedback is delivered.

However, it is not just HPC technology that needs to be considered here, as the data sizes involved mean that underlying data science and visualisation techniques must also be addressed. For instance, the VESTEC project is also researching novel data reduction techniques that support the identification of topologically relevant features from individual simulation step data. This relies on sampling raw data from topological proxies and then isolating the most representative members from the ensembles and estimating the probabilities of the appearance of specific features from this. Furthermore, graphical techniques such as in-situ ray tracing, must also be further developed to support scalable image generation and fast access to full-resolution simulation state.

\subsection{Dynamic ensemble simulations}
Previous research around using HPC for urgent decision making has identified how important it is for simulations to be started quickly, rather than languishing in the batch queue system \cite{startquickly}. Whilst using a dedicated machine for urgent decision making might seem like an obvious choice, the relative rarity of these disasters means that such a machine could be lightly used. Furthermore, the ability to run very large, high resolution jobs, is often desirable when it comes to urgent decision workloads, and as such the use of existing HPC resources is attractive. A common approach is to leverage queue priorities, such that jobs submitted to specific queues will run first or even terminate currently running jobs. However, this approach requires significant policy agreement with the individual HPC machine operators, and the execution of urgent jobs will inevitably cause disruption. This might not matter too much with previous urgent decision approaches, which involved running a small number of jobs, but our approach involves running ensemble simulations which can be started unpredictably based on sensor data streaming in over time, meaning that many runs over a not insignificant time period is likely. 

In VESTEC we have adopted the idea of a central VESTEC server federating over many HPC machines as per Figure \ref{fig:interactivity}. This system periodically tracks the utilisation of each HPC machine and, based upon this knowledge and the specifics of the job, will submit to which ever machine is deemed most appropriate. Federation is entirely hidden from the end-user and, whilst specific HPC machines might still employ further measures such as high priority queue which the VESTEC system can be aware of, the load will be spread out across multiple supercomputers if a significant number of jobs need to be executed. Being a European project, it is our vision that all the major supercomputers of Europe would eventually sign up to such a scheme, although for the purposes of the VESTEC research project we are federating over a smaller number of HPC machines to develop and prove the underlying concepts.

This approach solves another challenge in using HPC machines for urgent decision making, which is that of resilience. Most general purpose HPC machines simply do not operate with the guaranteed level of service availability required for life saving workloads. Even those that currently do, such as the Met Office, are fairly crude in their approach of relying on a hot backup machine which will clear itself of jobs and run the main workload if a failure is detected in the other machine. By contrast, in our approach the federator tracks which jobs are submitted to which machine, and if a supercomputer fails then not only will no more jobs be submitted to it, but also any queued or running jobs will be automatically resubmitted elsewhere. All of this is transparent to the user and more generally a single job might also be submitted to more than one machine speculatively, either to guard against machine failure or to simply hedge bets about which system will run the job first. 

Of course there are still challenges involved and an open question is whether this approach of polling machine status provides a level of control is fine grained enough to fully support the execution of codes within a bounded time frame. For instance, a major question which we are looking to answer in this project is whether it will be possible to collect enough data and make accurate enough predictions around machine state and queue times to provide this bounded execution time, or will some combination of special priority queues still be required in the most critical of situations?

Our federation approach does not get us entirely round the challenges of machine policy either. For instance, many operators of supercomputers have fairly strict policies around accounts and who may execute what. In some cases it might not be possible to have a single VESTEC user per system as this could be seen as account sharing which many systems disallow. Instead, individual users might still need their own accounts on the machines and their credentials then provided to VESTEC system for accessing the HPC machines on their behalf. Whilst this is not a major blocker in the context of a research project, in order to roll this approach out across very many machines, such as all supercomputers in Europe which is our vision, then such challenges must be considered and solved.

\subsection{Supporting an interactive workload}
When it comes to an interactive workload there are two concerns to bear in mind. Firstly, new data arriving from sensors must be processed in some manner, either by performing preliminary analysis or feeding it directly into simulation runs. Secondly, it is desirable to chain the running of simulation codes together, for instance, the execution of a weather model first generates a forecast which is then used by a simulation codes such as Wildfire Analyst, as details of the wind significantly impact how forest fires progress. Both of these requirements need to be handled in a way that requires minimal intervention from the already busy front line responders. 

\begin{figure}
\centering
\includegraphics[scale=0.4]{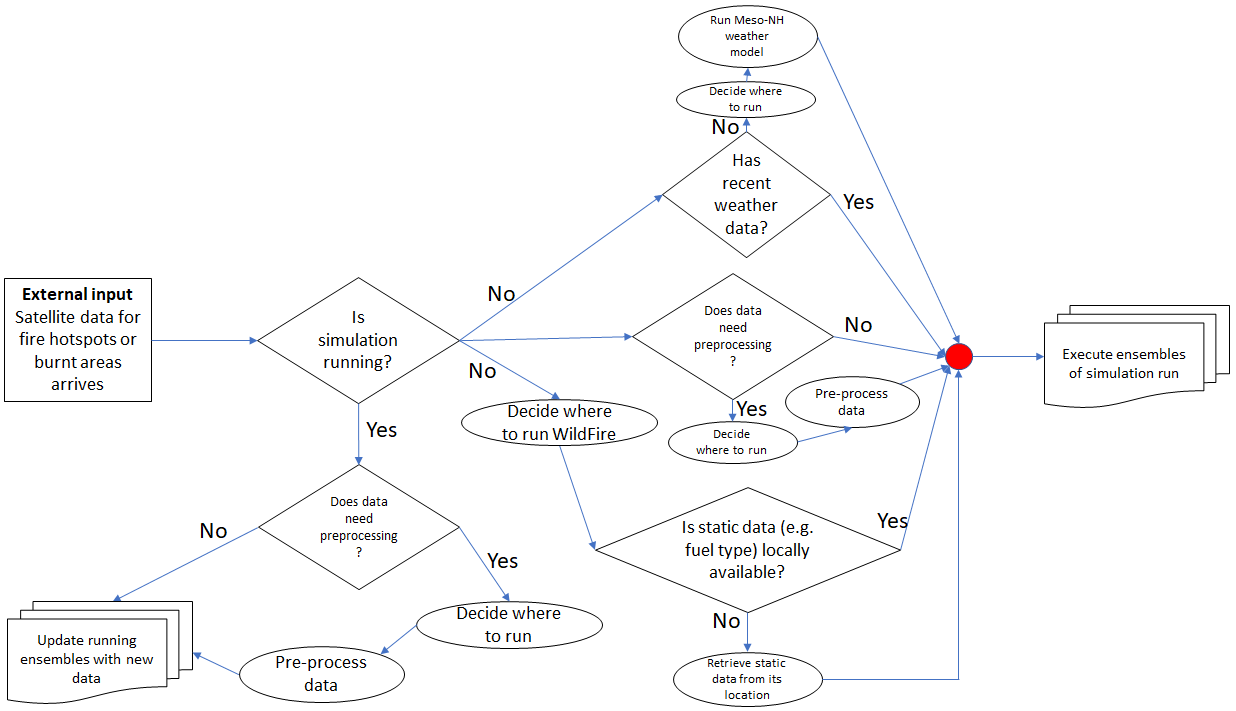}
\caption{Example workflow for executing or updating Wildfire Analyst in response to the arrival of new data}
\label{fig:workflow}
\end{figure}

We have identified that a workflow approach is appropriate here, where specific activities are performed and their status tracked. Once activities have completed then their results can be used to drive further activities as per a set of predefined rules. Figure \ref{fig:workflow} illustrates an example workflow for the forest fire use-case, where the arrival of new data will either update already running ensemble simulations, or start new ones. This is further complicated by the fact that the data might require pre-processing and new instances of the high resolution weather model might need to be executed too, and the weather forecast results provided to new instances of Wildfire Analyst. 

Whilst there are very many workflow technologies available, arguably the two most mature and ubiquitous are the Common Workflow Language (CWL) \cite{cwl} and Apache Taverna \cite{taverna}. In fact CWL is not a specific technology per-se, but instead a standard for describing workflows which numerous projects have implemented. By contrast, Apache Taverna provides its own workflow description language and ecosystem of tools such as GUIs for designing workflows. These technologies have not grown from an HPC perspective and as such neither is perfectly suited to our needs. When one considers the workflow in Figure \ref{fig:workflow}, they can observe a number of conditionals based on the state of the system or data. However, CWL does not support conditional branching in the workflows directly, so its ability to fully describe the workflows of our use-cases is limited. Apache Taverna by contrast does have some support for conditionals in the workflow, although this is not a commonly used feature, but Taverna is very heavy weight and at the current time of writing uses a description interface which is currently non-standard and in flux as the technology matures and moves towards full acceptance as an Apache project.

Therefore the approach we have adopted in the VESTEC project is to use CWL, due to its standardised nature, but only use this to describe each individual activity of the workflow and connect these using our own bespoke implementation in the VESTEC system. Whilst this isn't ideal, given the current state of these technologies we believe it is the best work around and potentially as these become more mature for HPC then we can move to standardising the links between activities. 

\section{The role of the cloud}
\label{sec:cloud}
From the discussions in Section \ref{sec:role} it might seem to the reader that many of the requirements for fusing real-time data with HPC for urgent decision making overlap with those of cloud computing. Hence a natural question is whether the adoption of some public cloud, such as Azure or AWS, could be appropriate here. One could go even further and argue that much of the work we are doing on the VESTEC system duplicates functionality already provided by the cloud, for instance supporting elastic compute, and as such a cloud-first solution could be more appropriate.

However, the cloud is not a silver bullet and there are specific limitations that impact its overall suitability for this kind of high performance work load. One such limitation is the fact that the user must set up the entirely of the infrastructure themselves and it is very easy to make decisions that, later down the line, significantly limit performance. Whilst the compute power of the cloud hardware and the interconnect is of high performance, selecting and configuring an appropriate file system can be a major issue. As such, it can be challenging to obtain high performance from codes that require significant IO, which is common place in HPC and involves all three of our VESTEC use-cases. Whilst high performance file-systems are often available on the cloud, these must be setup and configured by the user which is complex and beyond the capabilities of many users and developers. The common choice is to select a more common parallel file-system technology, such as NFS, and whilst this is still a non-trivial task to set up and configure, it is at-least doable in a realistic time frame.

\begin{figure}
\centering
\includegraphics[scale=0.75]{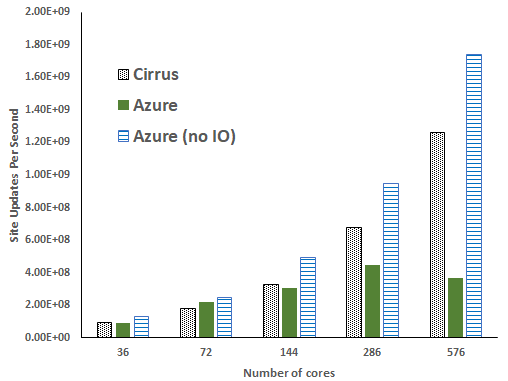}
\caption{Performance comparison, in Site Updates Per Second (higher is better) of HemeLB running on Cirrus, Azure and Azure with no IO}
\label{fig:cloud-performance}
\end{figure}

Figure \ref{fig:cloud-performance} illustrates a performance comparison for HemeLB \cite{hemelb}, a CFD code based on the lattice Boltzmann method, on Cirrus which is a HPE/SGI 8600 HPC system (36 Broadwell cores per node, connected via InfiniBand), and Azure (16 Haswell cores per node, connected via InfiniBand.) In this experiment we have picked a domain based on a 3d rotational angiography scan of an anonymous patient with a cerebral aneurysm being treated at the National Hospital for Neurology and Neurosurgery in London, UK. The simulations are concerned with modelling the region around the bifurcation of the internal carotid artery's bifurcation into the anterior cerebral artery and the middle cerebral artery and the code is configured to use a block size of 8, block counts of 13, 26, and 27, and 650492 total fluid sites. This code is not specifically part of one of the VESTEC use-cases, but we believe that this application area would be applicable to using HPC for urgent decision making in the future.

Whilst the processor technologies between Cirrus and Azure are different, we can still observe patterns here and draw some conclusions. The \emph{Cirrus} and \emph{Azure} runs are executing with IO in HemeLB enabled, whereas the \emph{Azure (no IO)} runs have had IO disabled at the application level. This last configuration, over Azure with no IO, is the fastest and this is because of the absence of IO, which adds some overhead to the code regardless. However it can be seen that when enabling IO on Azure the performance very significantly decreases, 4.7 times slower at 576 cores. The performance on Cirrus with IO enabled is 3.4 times faster than that of Azure with IO enabled. Whilst the Xeon processors are the next generation on Cirrus (Broadwell vs Haswell), we are still using the exact same number of cores and when profiling the code in more detail we found that the vast majority of this time difference was due to the overhead of IO on the cloud.

Another benefit of the cloud is that of elasticity, where new VMs can be spun up quickly rather than having to wait in a batch queue as per HPC machines. However, clouds are not always as elastic as they might initially seem and prior resource requests often have to be made ahead of time for significant amounts of compute. When one considers that the cloud companies must still provision the hardware and ensure that there is enough resource, then this isn't hugely surprising. It does however mean that requesting a very large amount of compute, very infrequently and unpredictably, as would be the case with urgent decision making for disaster response, is not what the cloud has been designed for.

\begin{figure}
\centering
\includegraphics[scale=0.75]{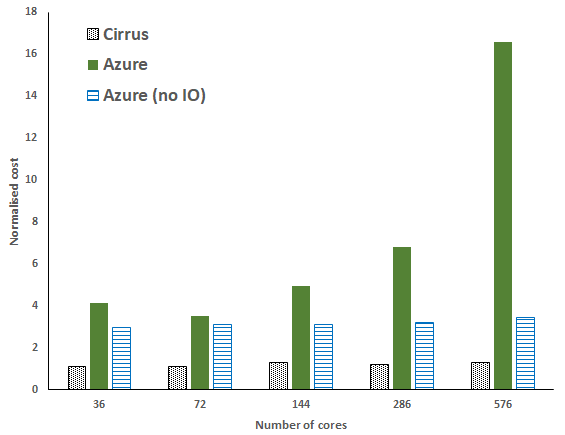}
\caption{Normalised cost (lower is better) comparison of running HemeLB running on Cirrus, Azure and Azure with no IO}
\label{fig:cloud-cost}
\end{figure}

There is also the question of cost and Figure \ref{fig:cloud-cost} illustrates a cost comparison between HemeLB running on Cirrus, Azure and Azure with no IO. The cost has been normalised, so we can explore the different configurations from a cost perspective irrespective of performance. It is important to note that the costs quoted here are the \emph{full} cost for Cirrus, including support, but Azure configurations only depict the cost of the VMs. Additional activities on the cloud after the job has finished, such as data storage, ingress and egress will incur additional charges. It can be seen that, running on Cirrus is consistently the cheapest option, being 2.7 times cheaper than Azure without any IO and over 12 times cheaper than Azure with IO at 576 cores. The most costly configuration is that of Azure with IO enabled and this is because of the additional file server support VMs that must be stood up. 

Whilst the numbers in Figures \ref{fig:cloud-performance} and \ref{fig:cloud-cost} illustrate the limitations of the cloud for HPC workloads, the cloud does have a number of features which could be of benefit in the use of computing for disaster response. Firstly, a cloud system could be one of our target machines, with the VESTEC system then federating over this exactly like the HPC machines. Secondly, we believe that some technologies which are closely tied with the cloud, such as object stores, will be very applicable to the technology solution employed by VESTEC.

In fact, in a way, the VESTEC system can be thought of a private cloud, with machines that subscribe to this then providing the compute resource and our VESTEC control system the overall marshaller. Whilst this does look different to public clouds such as Azure and AWS, there is an overlap of requirements and features, and whilst the implementation of these will by necessity be different, it could be said that the ability to provide the flexibility of the cloud but within the HPC space, is highly desirable for disaster response.

\section{Conclusions}
\label{sec:conclusions}
The use of HPC for urgent decision making is an exciting new domain for the super-computing community. However there are numerous challenges and barriers that need to be overcome in order for this to be a success, not least because  HPC machines, which traditionally favour throughput over individual job latency, are not set up for this sort of workload. In this paper we have identified three major challenges around interactivity that need to be addressed in order to successfully use HPC for disaster response; interacting with running simulations, dynamic ensemble control, and supporting an interactive workload. 

The role of interactivity is crucial here and, whilst no existing technologies are absolutely perfect, with further enhancements they can be made to work together but the devil is in the detail in terms of how one actually achieves this. It is clear that this effort requires expertise from across numerous domains, from traditional simulation, to data analysis, to visualisation techniques, to real-time computing, all these different components must be considered if we are to transform disaster response.

We are currently building the VESTEC system and exploring the hypothesis that by federating across multiple HPC machines, one addresses the limitations of HPC for this type of workload. Whilst it involves significant effort to further develop the tools, techniques and technologies, the potential payoff for the HPC and disaster response communities is significant if we are successful.

\section{Acknowledgements}
This work was funded under the EU FET VESTEC H2020 project, grant agreement number 800904.

%
%
%
%

\end{document}